\documentclass[aps,prb,reprint,secnumarabic,amsmath,amssymb,superscriptaddress,floatfix,eqsecnum]{revtex4-2}

\usepackage{amsmath}
\usepackage{amssymb}
\usepackage{amsfonts}
\usepackage{amsthm}
\usepackage{mathtools} % e.g., \coloneqq
\usepackage{bm}        % bold math symbols
\usepackage{bbm}       % e.g., \mathbbm{1}
\usepackage{slashed}
\usepackage{cancel}
\usepackage{cases}

\usepackage{graphicx}  % use PDF/PNG/JPG figures with pdflatex
\usepackage{xcolor}
\usepackage{multirow}
\usepackage{framed}
\usepackage{flafter}
\usepackage{float}

\usepackage{url}
\usepackage[colorlinks=true,citecolor=blue,linkcolor=blue,urlcolor=blue]{hyperref}

\begin{document}

\title{
Ground-state phase diagram of the spin-1/2 Heisenberg zigzag ladder with anisotropic $\Gamma$ term revisited
}

\author{Toshiya Hikihara}
\affiliation{Graduate School of Science and Technology, Gunma University, Kiryu, Gunma 376-8515, Japan}
\author{Nobuya Maeshima}
\affiliation{Center for Computational Sciences, University of Tsukuba, Tsukuba, Ibaraki 305-8577, Japan}
\affiliation{Institute of Pure and Applied Sciences, University of Tsukuba, Tsukuba, Ibaraki 305-8573, Japan}
\author{Kouichi Okunishi}
\affiliation{Department of Physics, Graduate School of Science, Osaka Metropolitan University, Osaka 558-8585, Japan}
\affiliation{Nambu Yoichiro Institute of Theoretical and Experimental Physics (NITEP), Osaka Metropolitan University, 3-3-138 Sugimoto, Osaka 558-8585, Japan}

\date{\today}

\begin{abstract}
We study the spin-1/2 Heisenberg antiferromagnet with an anisotropic $\Gamma$ term on a zigzag ladder, which can be exactly mapped to the spin-1/2 XYZ zigzag ladder by a spin rotation.
This model has recently attracted considerable interest as an effective model for the quasi-one-dimensional magnetic semiconductor YbCuS$_2$.
For the model, the ground-state phase diagram was previously proposed by Saito and Hotta [Phys.\ Rev.\ Lett. {\bf 132}, 166701 (2024); Phys.\ Rev.\ B {\bf 110}, 024409 (2024)].
We systematically reexamine the phase diagram through complementary numerical and analytical approaches, including exact diagonalization, the density-matrix renormalization group method, and low-energy effective theories.
We thereby demonstrate that when the anisotropic $\Gamma$ term is present, the model exhibits the N\'{e}el, dimer, ferromagnetic, and double-N\'{e}el phases, all of which have a finite excitation gap.
The gapless Tomonaga-Luttinger-liquid (TLL) phase appears only in the isotropic case where the $\Gamma$ term is absent.
We also confirm that the transitions between the N\'{e}el and dimer phases and between the N\'{e}el and ferromagnetic phases are continuous transitions governed by Gaussian criticality with central charge $c=1$.
The TLL parameter characterizing the critical states at the Gaussian transitions is estimated numerically.
\end{abstract}

\maketitle

\section{Introduction}\label{sec:introduction}

Low-dimensional frustrated quantum spin systems have long been the subject of extensive research as a platform for exploring the effect of strong quantum fluctuations in interacting many-body systems.\cite{SchollwockRFB2004}
Frustration arising from competing interactions suppresses conventional magnetic ordering and realizes a variety of novel quantum states.
A typical example is the one-dimensional (1D) spin-1/2 Heisenberg model on a zigzag-ladder lattice, which exhibits the dimerized ground states accompanied by spontaneous translational symmetry breaking.\cite{MajumdarG1969,Majumdar1970}
When the spin-rotational symmetry is further reduced by anisotropic interactions or an external magnetic field, the interplay between the frustration and anisotropy can give rise to various exotic phenomena, such as the vector-chirality ordering\cite{NersesyanGE1998,HikiharaKK2001,LecheminantJA2001,KolezhukV2005,Okunishi2008,McCullochKKKSK2008,HikiharaMFK2010,FurukawaSO2010,FurukawaSOF2012} and the formation of magnon-bound states.\cite{Tonegawa1970,OkunishiT2003,OkunishiT2003A,HMeisnerHV2006,KeckeMF2007,HikiharaKMF2008,SudanLL2009,DmitrievK2009,HMeisnerMK2009,HikiharaMFK2010}

Such interest in low-dimensional quantum magnets has recently been stimulated by the synthesis of Yb-based semiconductor YbCuS$_2$, in which Yb$^{3+}$ ions form a spin-1/2 zigzag ladder.\cite{Ohmagari2020}
Thermodynamic and magnetic measurements have revealed that the compound undergoes a magnetic transition at $T_N = 0.95$K.\cite{Ohmagari2020}
Subsequent $^{63/65}$Cu-nuclear magnetic resonance (NMR) and nuclear quadrupole resonance (NQR) have suggested the presence of exotic gapless excitations.\cite{HoriKKIMYOO2023}
$^{63/65}$Cu-NQR measurements under hydrostatic pressure\cite{HoriMKISO2025}, $^{63/65}$Cu-NMR measurements under magnetic fields\cite{HoriKIOO2025}, and the neutron diffraction experiments\cite{Onimaru2025} have also been conducted.

For YbCuS$_2$, a spin-1/2 Heisenberg zigzag ladder with an anisotropic $\Gamma$ term has been proposed as a microscopic model to describe its magnetic properties.\cite{SaitoNH2024}
The model Hamiltonian is given by
\begin{align}
\mathcal{H}
&= \sum_{\rho=1,2} \sum_{j} 
\left[ J_\rho {\bm s}_j \cdot {\bm s}_{j+\rho} 
+ \Gamma_\rho (s^x_j s^y_{j+\rho} + s^y_j s^x_{j+\rho} ) \right],
\label{eq:Ham_gamma}
\end{align}
where ${\bm s}_j$ is a spin-1/2 operator at the $j$th site.
$J_1$ and $J_2$ are respectively the exchange interaction constants of the nearest- and next-nearest-neighbor interactions, which are expected to be comparable, $J_1 \sim J_2$, for YbCuS$_2$.
Furthermore, the interaction constants for the $\Gamma$ term, $\Gamma_\rho$, ($\rho=1,2$) are expected to be smaller than $J_\rho$ by one or two orders of magnitude.
We set
\begin{align}
\gamma \equiv \frac{\Gamma_1}{J_1} = \frac{\Gamma_2}{J_2}
\label{eq:def_gamma}
\end{align}
and explore the model (\ref{eq:Ham_gamma}) in the $J_2/J_1$-$\gamma$ parameter plane.

The model (\ref{eq:Ham_gamma}) is intriguing from a theoretical viewpoint as well.
A notable point is that, under a global spin rotation by $\pi/4$ around the $s^z$ axis, the model (\ref{eq:Ham_gamma}) with Eq.\ (\ref{eq:def_gamma}) maps to the 1D $J_1$-$J_2$ XYZ model,\cite{SaitoH2024PRL}
\begin{align}
\mathcal{H}
&= \sum_{\rho=1,2} \sum_{j} J_\rho \left[
(1 - \gamma) S^x_j S^x_{j+\rho}
+ (1 + \gamma) S^y_j S^y_{j+\rho} \right.
\nonumber \\
&\left.~~~~~~~~~~~~~~~~
+ S^z_j S^z_{j+\rho} \right],
\label{eq:Ham_XYZ}
\end{align}
where
\begin{align}
S^x_j &= \frac{1}{\sqrt{2}} \left( s^x_j - s^y_j \right),~~
S^y_j  = \frac{1}{\sqrt{2}} \left( s^x_j + s^y_j \right),~~
S^z_j  = s^z_j.
\label{eq:s_to_S}
\end{align}
This $J_1$-$J_2$ XYZ model has been studied for various parameter regimes,\cite{MudryFMH2019,JiangM2019,RobertsJM2019,HuangLYMX2019,XuZGG2021} and is known to host a variety of phases including the dimer phase and several types of magnetically ordered phases.
In particular, the direct continuous phase transition between the N\'{e}el and dimer phases has attracted renewed interest in the context of the 1D analog of deconfined quantum criticality.\cite{MudryFMH2019,JiangM2019,RobertsJM2019,HuangLYMX2019}

In this work, we investigate the ground-state properties of the model (\ref{eq:Ham_gamma}) with Eq.\ (\ref{eq:def_gamma}), or equivalently, the model (\ref{eq:Ham_XYZ}).
For the models, previous research of Refs.\ \cite{SaitoH2024PRL,SaitoH2024PRB} has proposed a rich ground-state phase diagram and demonstrated interesting results including the presence of a multicritical point with degenerate ground states whose exact wave functions are expressed in a matrix-product form.
We revisit the models using complementary approaches such as the low-energy effective theories and numerical techniques, and determine the ground-state phase diagram for the parameter regime of $0 \le J_2/J_1 \le 1$ and $\gamma \ge 0$.
The phase diagram contains the N\'{e}el, dimer, ferromagnetic (FM), and double-N\'{e}el (DN) phases for $\gamma > 0$, and the Tomonaga-Luttinger-liquid (TLL) and dimer phases for $\gamma=0$.
We also demonstrate that a direct continuous transition occurs between the N\'{e}el and dimer phases.
The transition between the N\'{e}el and FM phases is also continuous.
These two continuous transitions are the Gaussian transitions with a central charge $c=1$, and we numerically estimate the TLL parameter along the transition lines.

The rest of this paper is organized as follows.
Section\ \ref{sec:method} describes the numerical methods employed and the quantities analyzed in this work.
Section\ \ref{sec:result} presents the results for the ground state.
We first provide an overview of the ground-state phase diagram and characterize the phases appearing therein in Sec.\ \ref{subsec:gsPD}.
Then, Sec.\ \ref{subsec:Neel-Dim}, \ref{subsec:Neel-FM}, and \ref{subsec:DN} are devoted respectively to the transition between the N\'{e}el and dimer phases, the transition between the N\'{e}el and FM phases, and the transitions to the DN phase.
Section\ \ref{sec:conclusion} presents our conclusions.

\section{Methods}\label{sec:method}

We have determined the ground-state phase diagram of the model (\ref{eq:Ham_gamma}) with Eq.\ (\ref{eq:def_gamma}), or the equivalent model (\ref{eq:Ham_XYZ}), through numerical calculations using the exact diagonalization (ED) and density-matrix renormalization group (DMRG) methods, combined with results from low-energy effective theories.
In this section, we summarize the details of the numerical calculations.
We note that we have performed our analyses mainly for the model (\ref{eq:Ham_XYZ}), which is represented in terms of the spins ${\bm S}_j$; the results can be immediately expressed in terms of the spins ${\bm s}_j$ in the model (\ref{eq:Ham_gamma}) through Eq.\ (\ref{eq:s_to_S}).

We have explored the energy spectrum of the model (\ref{eq:Ham_gamma}) using the ED method.
The ED calculations were performed for the system under the periodic boundary conditions with up to $L=28$ spins.
In addition to our in-house ED code, we employed the QuSpin and XDiag packages \cite{QuSpin2017,xdiag,xdiag-0.4} to calculate the low-lying energy levels and the corresponding excitation gaps,
\begin{align}
\Delta_n(L) &= E_n(L) - E_0(L),
\label{eq:Energy_Gap}
\end{align}
where $E_0(L)$ is the ground-state energy and $E_n(L)$ with $n \ge 1$ is the energy of the $n$th excited state.

The DMRG calculations have been performed for the model (\ref{eq:Ham_XYZ}) with $L$ spins under the open boundary conditions.
In the calculations of the magnetic order parameters, we may add to the model (\ref{eq:Ham_XYZ}) the boundary Zeeman term,
\begin{align}
\mathcal{H}_h &= - h^x_1 S^x_1 - h^x_L S^x_L - h^y_1 S^y_1 - h^y_L S^y_L,
\label{eq:Ham_boundarifield}
\end{align}
to lift the degeneracy of the magnetically-ordered states.
The DMRG calculation was performed typically for the system size up to $L=192$, while the calculation for $L=256$ was also done in some cases.
The bond dimension was up to $\chi=160$.
The truncation error, the average of the density-matrix weight of discarded bases over the final DMRG sweep, was typically less than $1 \times 10^{-10}$, and less than $4 \times 10^{-9}$ even at the worst case.

Using the DMRG method, we have calculated the local spin polarizations $\langle S^x_j \rangle$ and $\langle S^y_j \rangle$, and the nearest-neighbor correlation functions $\langle S^x_j S^x_{j+1}\rangle$, $\langle S^y_j S^y_{j+1}\rangle$, and $\langle S^z_j S^z_{j+1}\rangle$, where $\langle \cdots \rangle$ denotes the ground-state expectation value.
In the magnetic phases expected to appear in the model (\ref{eq:Ham_XYZ}) for the parameter region considered, the magnetic orders with up to four-site period are anticipated.
Therefore, we have examined the following order parameters near the center of the open chain with $L$ spins;
\begin{align}
M^a_{\rm F} &= \frac{1}{2} 
\langle ( S^a_{\frac{L}{2}-1} + S^a_{\frac{L}{2}} + S^a_{\frac{L}{2}+1} + S^a_{\frac{L}{2}+2} ) \rangle,
\label{eq:MagOP_F} \\
M^a_{\rm N} &= \frac{1}{2} 
\langle ( S^a_{\frac{L}{2}-1} - S^a_{\frac{L}{2}} + S^a_{\frac{L}{2}+1} - S^a_{\frac{L}{2}+2} ) \rangle,
\label{eq:MagOP_N} \\
M^a_4       &= \frac{1}{2} 
\langle ( - S^a_{\frac{L}{2}-1} + S^a_{\frac{L}{2}} + S^a_{\frac{L}{2}+1} - S^a_{\frac{L}{2}+2} ) \rangle,
\label{eq:MagOP_4}
\end{align}
where $a=x,y$.
We have also analyzed the dimer order parameter,
\begin{align}
\mathcal{O}_{\rm D} &= \langle {\bm S}_{\frac{L}{2}} \cdot {\bm S}_{\frac{L}{2}+1} \rangle - \langle {\bm S}_{\frac{L}{2}-1} \cdot {\bm S}_{\frac{L}{2}} \rangle.
\end{align}

To investigate the nature of the critical state at the phase transitions, we have computed the entanglement entropy (EE) and the local energy density.
The EE examined is the one between the blocks of the left $l$ sites and the right $L-l$ sites,
\begin{align}
\mathcal{S}(l) &= - {\rm Tr}_l \hat{\rho}_l \ln \hat{\rho}_l,
\end{align}
where $\hat{\rho}_l$ is the reduced density matrix for the block of the left $l$ sites and the trace ${\rm Tr}_l$ is taken for the left block.
The local energy density is defined as,
\begin{align}
E_{\rm b}(j) &= J_1 e_{\rm b}(j, j+1)
\nonumber \\
&~~~+ \frac{1}{2} J_2 \left[ e_{\rm b}(j-1, j+1) + e_{\rm b}(j, j+2) \right],
\label{eq:bndE}
\end{align}
where 
\begin{align}
e_{\rm b}(j, j') 
&= (1-\gamma) \langle S^x_j S^x_{j'} \rangle
 + (1+\gamma) \langle S^y_j S^y_{j'} \rangle
 + \langle S^z_j S^z_{j'} \rangle.
\end{align}

\section{Results}\label{sec:result}
\subsection{Ground-State Phase Diagram}\label{subsec:gsPD}

The ground-state properties of the model (\ref{eq:Ham_gamma}) with Eq.\ (\ref{eq:def_gamma}) were investigated in Refs.\ \cite{SaitoH2024PRL,SaitoH2024PRB}, where the ground-state phase diagram including the dimer phase, a TLL phase, and various magnetically ordered phases was proposed.
It was also shown that there existed a multicritical point at $(J_2/J_1, \gamma)=(1/2, \sqrt{3})$, where the degenerate ground states were described by exact matrix-product wave functions.
We systematically reexamine the phase diagram using complementary approaches such as the low-energy effective theories and the numerical analyses of the ED and DMRG methods.

\begin{figure}
\centering
\includegraphics[width=70mm]{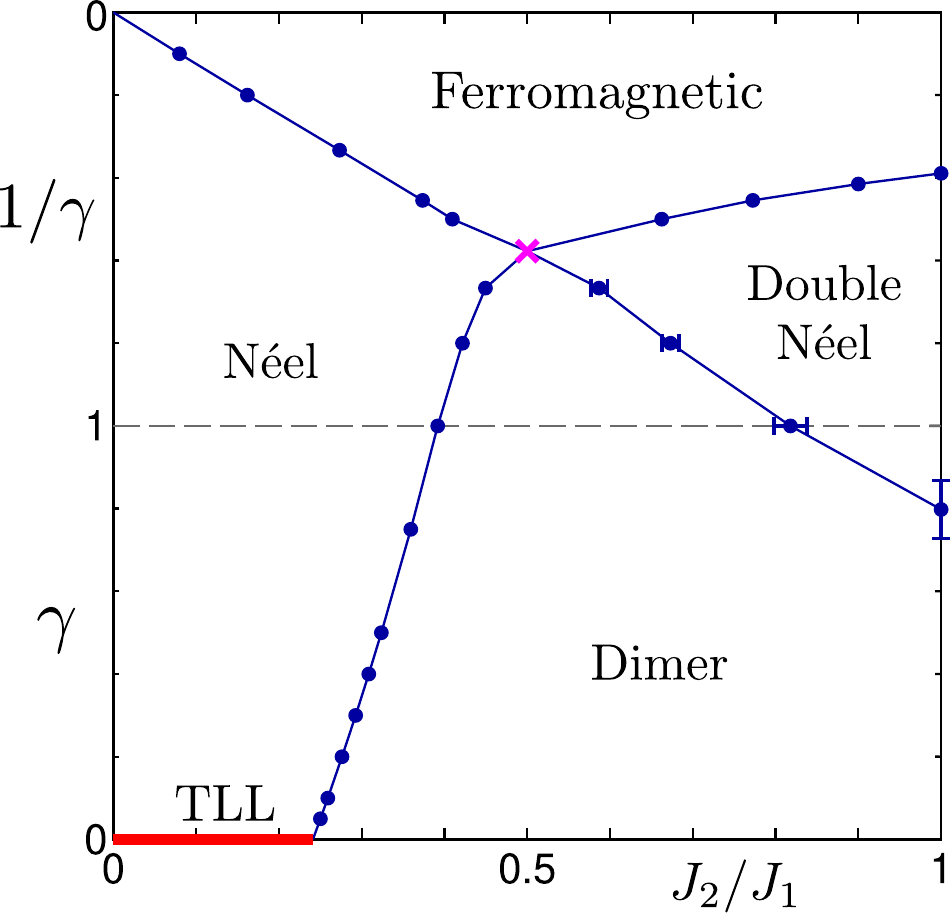}
\caption{
Ground-state phase diagram of the model (\ref{eq:Ham_gamma}) or (\ref{eq:Ham_XYZ}) in the parameter plane of $J_2/J_1$ and $\gamma = \Gamma_1/J_1 = \Gamma_2/J_2$.
The region for $\gamma > 1$ is shown as a function of $1/\gamma$.
The cross at $(J_2/J_1, \gamma)=(1/2, \sqrt{3})$ represents the multicritical point.\cite{SaitoH2024PRL}
Red line at $\gamma=0$ and $J_2/J_1 < 0.2411...$ represents the TLL phase.
}
\label{fig:GSPD}
\end{figure}

\begin{figure}
\centering
\includegraphics[width=78mm]{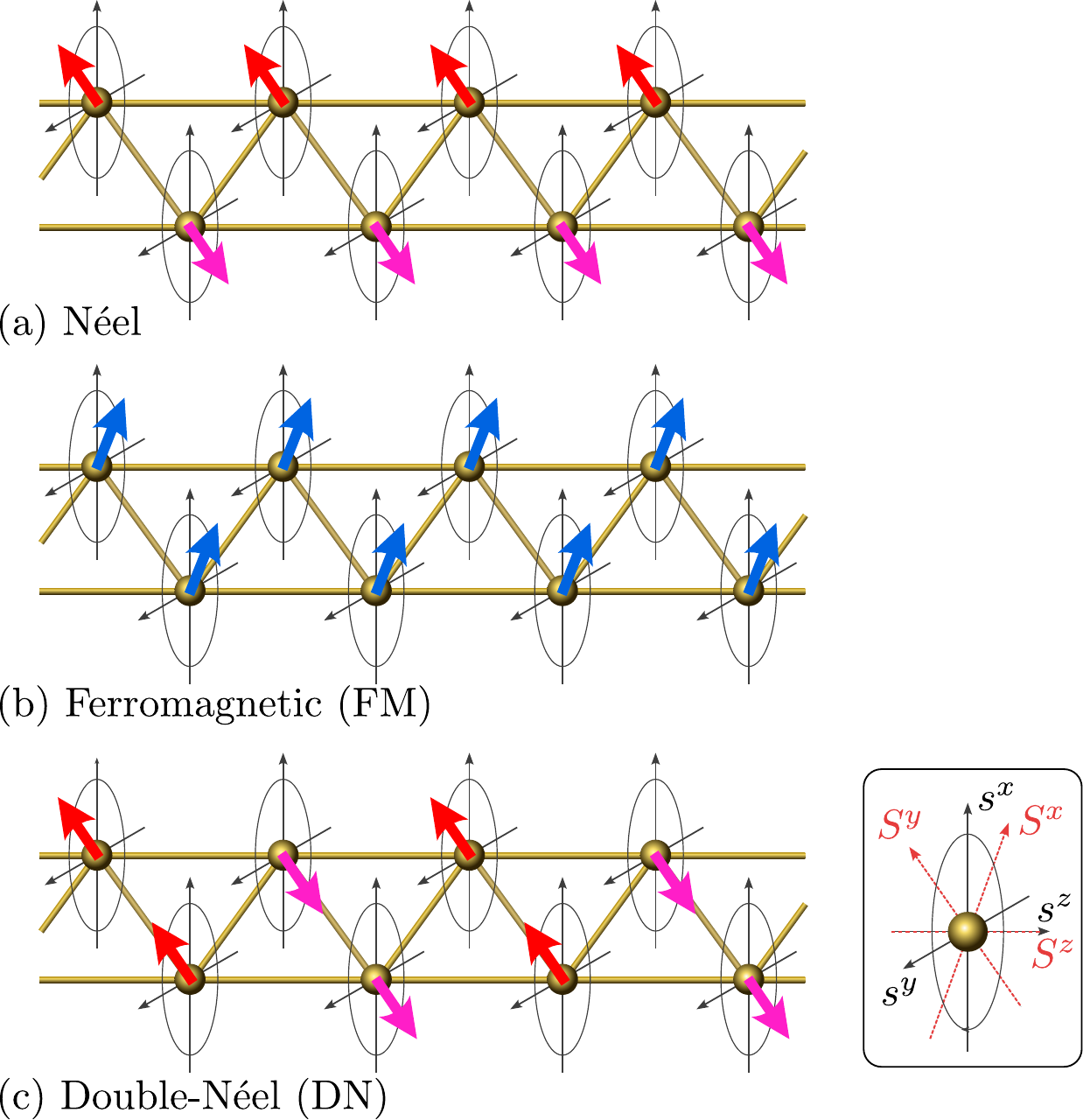}
\caption{
Schematic picture of the magnetic order in (a) the N\'{e}el phase, (b) the ferromagnetic (FM) phase, and (c) the double-N\'{e}el (DN) phase.
}
\label{fig:mag_order}
\end{figure}

Figure\ \ref{fig:GSPD} presents the ground-state phase diagram of the model (\ref{eq:Ham_gamma}) or (\ref{eq:Ham_XYZ}) obtained for $0 \le J_2/J_1 \le 1$ and $\gamma \equiv \frac{\Gamma_1}{J_1} = \frac{\Gamma_2}{J_2} \ge 0$.
Note that the sign of $\gamma$ can be reversed by a $\pi/2$ rotation of the spins around the $s^z=S^z$ axis, and the phase diagram for $\gamma < 0$ is the inversion of that for $\gamma > 0$ with appropriate changes in the direction of the magnetic orders.
The phase diagram contains five phases: 
\begin{itemize}
\item[(i)] {\it N\'{e}el phase}: 
The N\'{e}el phase appears in the region where $\gamma > 0$ and $J_2/J_1$ is small.
In this phase, the system exhibits the staggered magnetic long-range order along the $S^y$ direction for spins in the model (\ref{eq:Ham_XYZ}).
For spins in the model (\ref{eq:Ham_gamma}), the direction of the magnetic order corresponds to $(s^x, s^y, s^z) = (\pm 1, \pm 1,0)$.
In the thermodynamic limit, the ground states are doubly degenerate,\cite{SSB_footnote} and the excitation energy above the ground states is gapped.
\item[(ii)] {\it Dimer phase}: 
The dimer phase appears in the region for not too large $\gamma$ and intermediate $J_2/J_1$.
In this phase, the dimer order parameter is nonzero, $\mathcal{O}_{\rm D} \ne 0$, due to the spontaneous breaking of translational symmetry, and no magnetic order exists.
When $\gamma > 0$, the nematic order $s^x_j s^y_{j+1} + s^y_j s^x_{j+1} = S^y_j S^y_{j+1} - S^x_j S^x_{j+1}$ persists\cite{SaitoH2024PRL,SaitoH2024PRB}, reflecting the explicit anisotropy of the model Hamiltonians.
In the thermodynamic limit, the system possesses the doubly-degenerate ground states and gapped excitations above them.
\item[(iii)] {\it Ferromagnetic (FM) phase}: 
The FM phase appears in the region where $\gamma$ is large and $J_2/J_1$ takes an intermediate value.
The FM phase is characterized by a uniform magnetic long-range order along the $S^x$ direction.
In terms of the spins ${\bm s}_j$, the direction of the magnetic order is $(s^x, s^y, s^z) = (\pm 1, \mp 1,0)$.
The ground states are doubly degenerate and the excitations are gapped in the thermodynamic limit.
\item[(iv)] {\it Double-N\'{e}el (DN) phase}: 
The DN phase appears in the region where $\gamma$ is intermediate and $J_2/J_1$ is large.
The ground state of the DN phase can be viewed as consisting of two N\'{e}el-ordered chains, each formed by the $J_2$ couplings.
Accordingly, the DN phase is characterized by the magnetic long-range order with an up-up-down-down pattern along the $S^y$ direction [the direction of $(s^x, s^y, s^z) = (\pm 1, \pm 1,0)$].
In the thermodynamic limit, the ground states are four-fold degenerate, and the excitations above them are gapped.
\item[(v)] {\it Tomonaga-Luttinger liquid (TLL) phase}: 
The TLL phase appears on the SU(2)-symmetric parameter line at $\gamma=0$ and for $J_2/J_1 < 0.2411...$.\cite{Haldane1982,Haldane1982E,NomuraO1994,OkamotoN1992,Eggert1996}
The TLL phase is characterized by the absence of any long-range order, and instead, by quasi-long-range order in algebraically decaying spin correlations.
The system possesses the unique ground state and gapless excitations above it.
\end{itemize}
The spatial pattern of the magnetic orders in (i) N\'{e}el, (iii) FM, and (iv) DN phases, is depicted in Fig.\ \ref{fig:mag_order}.
Here, the $s^z=S^z$ axis is parallel to the legs of the zigzag ladder, the $s^x$ axis in the plane of the zigzag triangles, and the $s^y$ axis perpendicular to the plane.\cite{SaitoNH2024}

The transitions between the N\'{e}el and dimer phases and between the N\'{e}el and FM phases are continuous transitions described by the Gaussian theory with the central charge $c=1$.
The transition between the FM and DN phases is the first-order transition.
These transition lines and the dimer-DN transition line meet at the multicritical point at $(J_2/J_1, \gamma)=(1/2, \sqrt{3})$.

The most significant difference in our phase diagram from the one in Refs.\ \cite{SaitoH2024PRL,SaitoH2024PRB} is that the TLL phase does not appear in the region with $\gamma > 0$ where the model is anisotropic.
This absence of the TLL phase for $\gamma > 0$ can be deduced from the results of previous studies described below.

The primarily important point is that the model (\ref{eq:Ham_gamma}) or (\ref{eq:Ham_XYZ}) for $\gamma > 0$ has no continuous symmetries.
Therefore, the TLL phase does not appear unless a continuous symmetry at the low-energy limit emerges through some additional mechanism.
Indeed, the low-energy effective theories for the $J_1$-$J_2$ XYZ model [Eq.\ (\ref{eq:Ham_XYZ})] indicate that the model exhibits only gapped phases (except the critical point at phase transitions).\cite{MudryFMH2019,LeClairS2004}
By the standard bosonization procedure, the low-energy effective Hamiltonian for the model (\ref{eq:Ham_XYZ}) in terms of the bosonic fields $\{ \phi, \theta \}$ is derived as\cite{MudryFMH2019}
\begin{align}
\tilde{\mathcal{H}} &= \frac{v}{2} \int dx \left[ \frac{1}{\eta}(\partial_x \theta)^2 + \eta (\partial_x \phi)^2 \right]
\nonumber \\
&~~+ \lambda_\phi \int dx \cos(\sqrt{8\pi}\phi) 
+ \lambda_\theta \int dx \cos(\sqrt{8\pi}\theta),
\label{eq:Ham_XYZ_boson}
\end{align}
where $v$ is the spin velocity, $\eta$ is the TLL parameter, and $\lambda_\phi$ and $\lambda_\theta$ are coupling constants.
The spin operators are related to the bosonic fields as
\begin{align}
S^z_j &\simeq \frac{1}{\sqrt{2\pi}} \partial_x \phi(x) 
+ b_1 (-1)^j \sin(\sqrt{2\pi}\phi(x)) + \cdots, 
\label{eq:Sz_to_bosonfield} \\
S^+_j &\simeq e^{i\sqrt{2\pi}\theta(x)}
\left[ b_2 (-1)^j + b_3 \sin(\sqrt{2\pi}\phi(x))\right] + \cdots, 
\label{eq:S+_to_bosonfield}
\end{align}
where $b_1$, $b_2$, $b_3$ are real constants, and we have set the lattice spacing to unity so that the continuous coordinate $x$ is equal to $j$.
The term $\cos(\sqrt{8\pi} \theta)$ in the effective Hamiltonian (\ref{eq:Ham_XYZ_boson}) originates from the anisotropic term of $\gamma$,
\begin{align}
- J_\rho \gamma \left( S^x_j S^x_{j+\rho} - S^y_j S^y_{j+\rho} \right) 
&= -\frac{J_\rho \gamma}{2} \left( S^+_j S^+_{j+\rho} + S^-_j S^-_{j+\rho} \right) 
\nonumber \\
&\simeq - J_\rho \gamma b_2^2 (-1)^{\rho} \cos(\sqrt{8\pi}\theta).
\label{eq:Qxy_boson}
\end{align}

For the isotropic parameter line $\gamma=0$, the effective model (\ref{eq:Ham_XYZ_boson}) reduces to the single sine-Gordon model, and it has been established that the model exhibits the Berezinskii-Kosterlitz-Thouless transition between the TLL phase for small $J_2/J_1$ and the dimer phase for large $J_2/J_1$.\cite{Haldane1982,Haldane1982E,NomuraO1994}
The transition point is located at $J_2/J_1=0.2411...$.\cite{OkamotoN1992,Eggert1996}
Then, when the anisotropy of $\gamma > 0$ is introduced, the effective model becomes the double sine-Gordon model, and the weak-coupling RG analysis around the TLL-dimer transition point revealed that the model with $\gamma > 0$ realized either the gapped phase with the N\'{e}el order in $S^y$ operators or the gapped dimer phase;
The model can be gapless only at the transition between these gapped phases.\cite{MudryFMH2019,LeClairS2004,LecheminantGN2002}
Therefore, the TLL phase does not appear in an extended region of the parameter space for $\gamma > 0$.
At the transition between the N\'{e}el and dimer phases, it has been shown that the model (\ref{eq:Ham_XYZ}) exhibits the Gaussian criticality with the central charge $c=1$, which can be regarded as a 1D analog of deconfined quantum criticality.\cite{MudryFMH2019,JiangM2019,RobertsJM2019,HuangLYMX2019}

The absence of the TLL phase for the anisotropic case, $\gamma > 0$, is also indicated by an exact analysis for $J_2=0$.
For this simple anisotropic XYZ chain, the exact solution has revealed that the excitation spectrum exhibits a finite energy gap above the doubly degenerate ground states.
The energy gap in the thermodynamic limit corresponds to the bottom of the two-spinon continuum band, and is expressed in a closed form involving the complete elliptic integral and the Jacobi elliptic function.\cite{JohnsonJD1973}
A detailed derivation of the two-spinon gap $\Delta_{\rm 2sp}(\gamma)$ is described in Ref.\ \cite{MaeshimaOH_proc}.
In particular, the asymptotic form of the energy gap at the limit of small $\gamma$ is obtained as
\begin{align}
\Delta_{\rm 2sp}(\gamma) \simeq 4\pi(1+\gamma) \exp\left( - \frac{k}{\sqrt{\gamma}}\right),
\end{align}
where $k = 2.912 ...$.
Thus, for the model (\ref{eq:Ham_XYZ}) with $J_2=0$, one can conclude that the energy gap takes a finite value for $\gamma > 0$, although it is exponentially small for a small $\gamma$.

Based on the above results, we conclude that the TLL phase appears only on the isotropic parameter line $\gamma = 0$ with $J_2/J_1 < 0.2411...$.
For $\gamma > 0$, the phase diagram consists of three magnetic ordered phases and the dimer phase.
These four phases at $\gamma > 0$ all have a finite excitation gap, although the gap may be exponentially small for the region where $J_2/J_1$ and $\gamma$ are small.
We will discuss in detail the nature of the phase transitions in the following sections.

\subsection{Transition between N\'{e}el and dimer phases}\label{subsec:Neel-Dim}
As mentioned in the previous section, the N\'{e}el-dimer transition is described by the double sine-Gordon model (\ref{eq:Ham_XYZ_boson}) containing two competing cosine terms.\cite{MudryFMH2019}
For $J_2/J_1 < (J_2/J_1)_{\rm c}$, the term of $\cos(\sqrt{8\pi}\theta)$ wins the competition, and $\lambda_\theta$ flows to a large positive value, leading to the N\'{e}el phase with the staggered magnetic order along the $S^y$ direction.
For $J_2/J_1 > (J_2/J_1)_{\rm c}$, the term of $\cos(\sqrt{8\pi}\phi)$ dominates, and $\lambda_\phi$ flows to the negative strong limit, resulting in a gapped phase with the dimer long-range order.
At the transition point $J_2/J_1 = (J_2/J_1)_{\rm c}$, the system exhibits a criticality governed by the Gaussian model.
The critical value $(J_2/J_1)_{\rm c}$ depends on $\gamma$ and is determined numerically below.

To determine the transition line between the N\'{e}el and dimer phases, we have performed an analysis of the scaled gap $L \Delta_1(L)$ in the system under the periodic boundary conditions.
In the N\'{e}el and dimer phases, the level splitting $\Delta_1(L)$ [Eq.\ (\ref{eq:Energy_Gap})] between the ground state and the first excited state decreases with the system size $L$, and the two states constitute the doubly degenerate ground states in the thermodynamic limit $L \to \infty$.
If the system size $L$ is sufficiently large, the level splitting $\Delta_1(L)$ is expected to decay exponentially as $L$ increases, and the scaled gap $L\Delta_1(L)$ also decays to zero as $L \to \infty$.
At the critical point, the energy levels of the first excited states of the two phases cross, resulting in the cusp in $\Delta_1(L)$.
Furthermore, since $\Delta_1(L)$ decays as $\sim 1/L$, the scaled gap $L\Delta_1(L)$ converges to a finite value.
Therefore, the transition between the N\'{e}el and dimer phases is signaled by the cusp singularity in the scaled gap $L \Delta_1(L)$, whose height remains finite at $L \to \infty$.\cite{NomuraO1994}

\begin{figure}
\centering
\includegraphics[width=80mm]{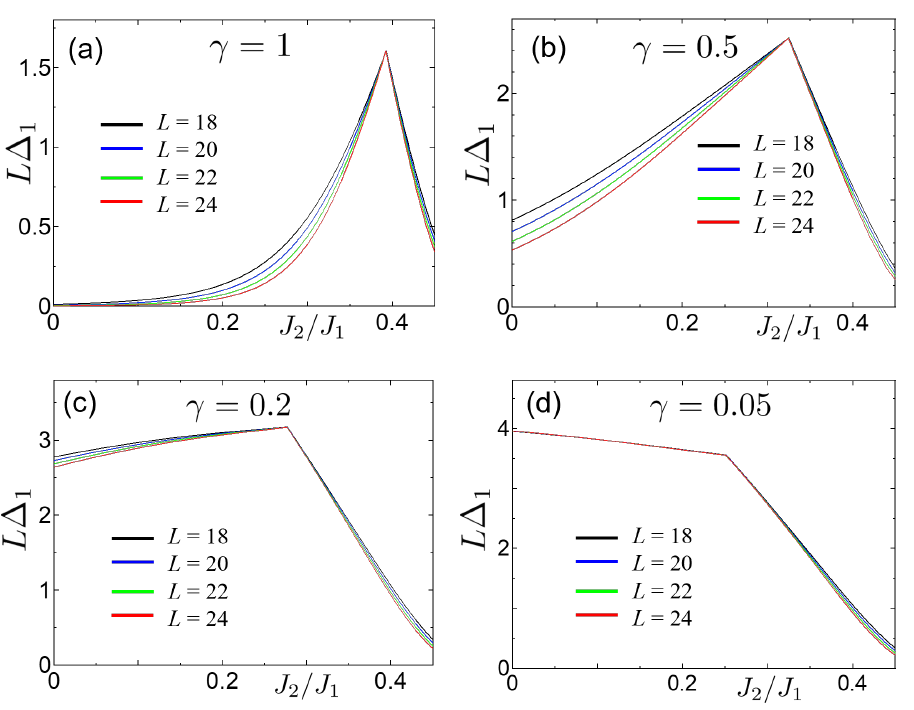}
\caption{
Scaled gap $L\Delta_1(L)$ as a function of $J_2/J_1$ along parameter lines with (a) $\gamma=1.0$, (b) $\gamma=0.5$, (c) $\gamma=0.2$, and (d) $\gamma=0.05$.
}
\label{fig:SGap_Neel-Dim}
\end{figure}

In Fig.\ \ref{fig:SGap_Neel-Dim}, we show the numerical data of the scaled gap $L \Delta_1(L)$ obtained by the ED method for the parameter lines with $\gamma=1.0, 0.5, 0.2, 0.05$ around the N\'{e}el-dimer transition as a function of $J_2/J_1$.
For $\gamma=1.0$, the scaled gap exhibits a clear cusp singularity at $J_2/J_1 \sim 0.39$.
The height of the cusp converges to a nearly constant value independent of $L$, suggesting that a quantum criticality with $\Delta_1(L) \sim 1/L$ is realized at the transition point.
On both sides of the transition point, the scaled gap decreases with $L$, indicating the rapid decay, being faster than $1/L$, of the level splitting $\Delta_1(L)$.
As $\gamma$ is smaller, the decay of the scaled gap in both the N\'{e}el and dimer phases is slower;
In the N\'{e}el phase for $\gamma=0.05$, the expected decay of the scaled gap is no longer observable.
This result suggests that the correlation length of the model in these regions of the gapped phases is so large that we can not observe the asymptotic behavior of $\Delta_1(L)$ in the system size $L \le 24$.
However, even for such small $\gamma$'s, the cusp singularity in $L\Delta_1(L)$ remains distinct, and one can confirm the level crossing of the first excited states at the transition.
We have estimated the N\'{e}el-dimer transition point for each finite $L$ from the position of the cusp in the scaled gap, and extrapolated them to $L \to \infty$ by fitting the data to a quadratic polynomial of $1/L$. 
The transition points thus obtained are plotted in the phase diagram in Fig.\ \ref{fig:GSPD}.

\begin{figure}
\centering
\includegraphics[width=75mm]{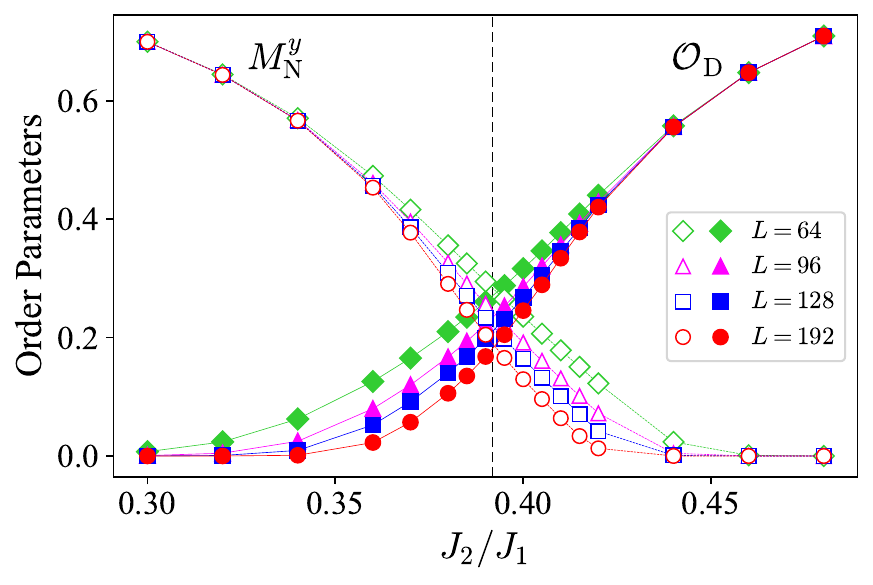}
\caption{
Order parameters as a function of $J_2/J_1$ along the parameter line with $\gamma=1.0$.
Open and solid symbols represent the N\'{e}el order parameter $M^y_{\rm N}$ and the dimer order parameter $\mathcal{O}_{\rm D}$, respectively.
Vertical dashed line represents the boundary between the N\'{e}el and dimer phases.
}
\label{fig:OP_Neel-Dim}
\end{figure}

To see the behavior of the order parameters around the N\'{e}el-dimer transition, we have performed the DMRG calculation.
Figure\ \ref{fig:OP_Neel-Dim} shows the N\'{e}el order parameter along the $S^y$ direction, $M^y_{\rm N}$, and the dimer order parameter $\mathcal{O}_{\rm D}$ for the parameter line of $\gamma=1.0$.
The N\'{e}el order parameter was calculated in the system with the boundary fields $h^y_1 = - h^y_L = 100J_1$ and $h^x_1 = h^x_L = 0$, while no boundary field was applied in the calculation of the dimer order parameter $\mathcal{O}_{\rm D}$.
The results indicate that the N\'{e}el-type magnetic long-range order along the $S^y$ direction emerges in the N\'{e}el phase while the dimer long-range order appears in the dimer phase.

\begin{figure}
\centering
\includegraphics[width=75mm]{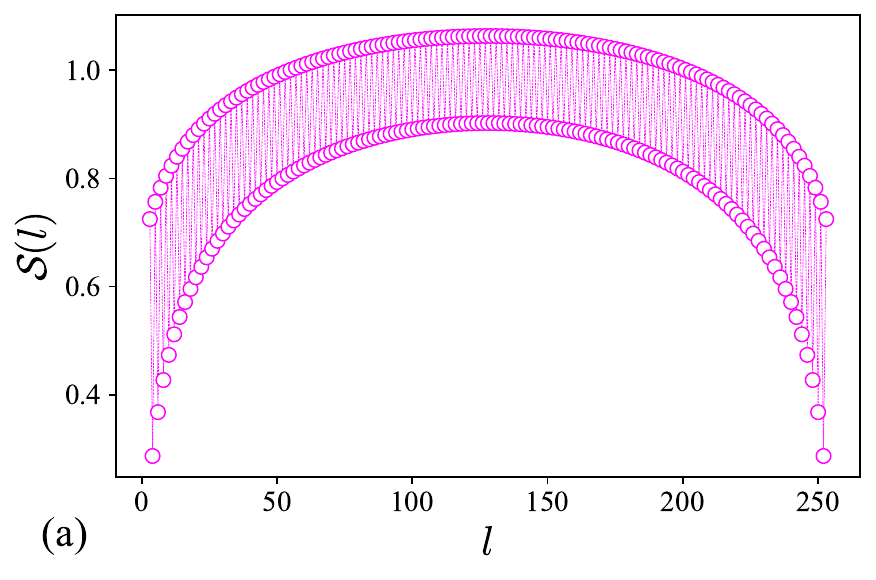}
\includegraphics[width=75mm]{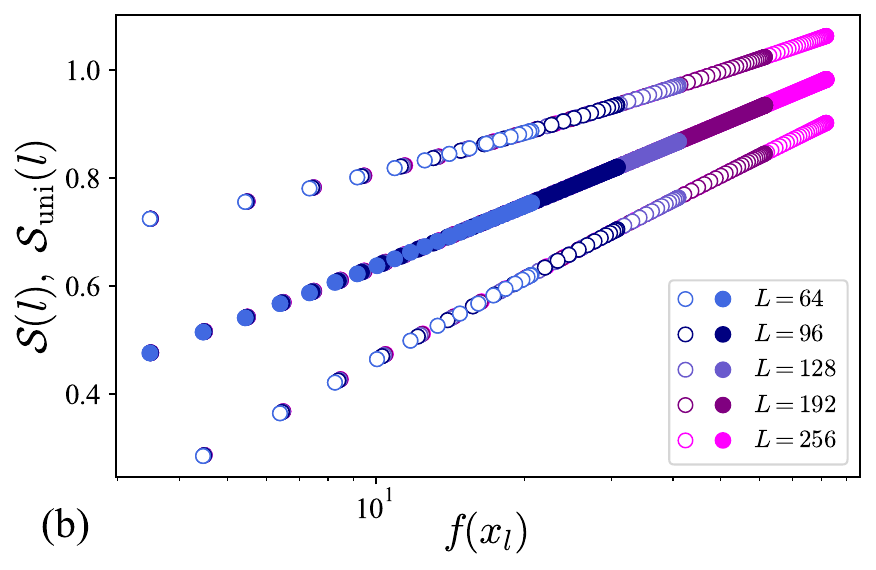}
\includegraphics[width=75mm]{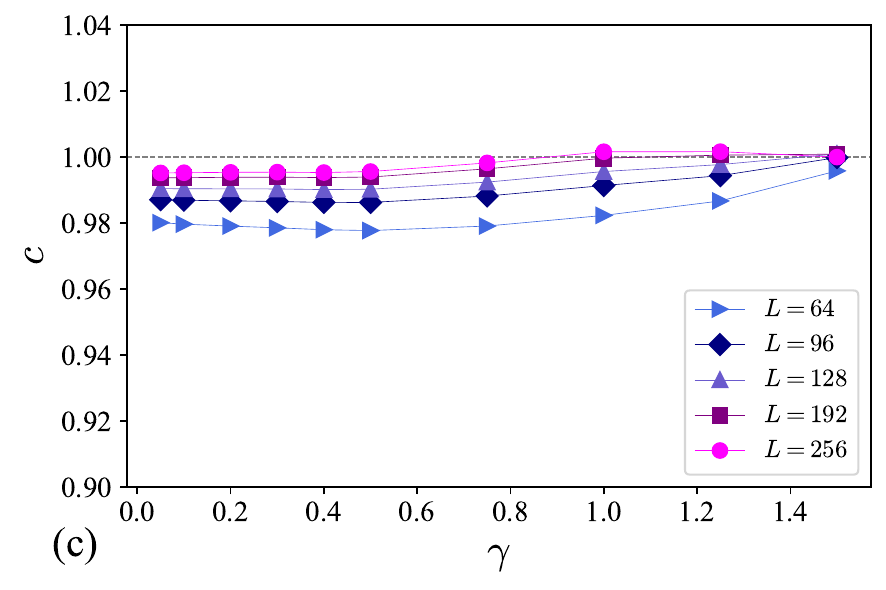}
\caption{
(a) Entanglement entropy $\mathcal{S}(l)$ of the system with $L=256$ spins at the N\'{e}el-dimer transition point for $\gamma=1.0$, plotted as a function of $l$.
(b) $\mathcal{S}(l)$ (open symbols) and its uniform part $\mathcal{S}_{\rm uni}(l)$ (solid symbols) for several $L$ at the N\'{e}el-dimer transition point for $\gamma=1.0$, plotted as a function of $f(x_l)$.
(c) The central charge $c$ at the N\'{e}el-dimer transition as a function of $\gamma$ for several $L$.
}
\label{fig:EEnt_Neel-Dim}
\end{figure}

Next, to investigate the properties of the critical state at the N\'{e}el-dimer transition, we have performed the DMRG calculation to obtain the EE $\mathcal{S}(l)$ at the transition points.
The calculation was done for the system under the open boundary conditions and without the boundary fields.
It is known that the EE in a 1D critical system under the open boundary conditions takes the following form,\cite{HolzheyLW1994,VidalLRK2003,CalabreseC2004,LaflorencieSCA2006,AffleckLS2009}
\begin{align}
\mathcal{S}(l) &= \frac{c}{6} \ln[f(x_l)] + \mathcal{S}_{\rm b,osc}(l) + \mathcal{S}_0,
\label{eq:EE_1Dform}
\end{align}
where $c$ is the central charge, $\mathcal{S}_0$ is a constant, and the function\begin{align}
f(x_l) &= \frac{L+1}{\pi}\sin\left( \frac{\pi x_l}{L+1}\right),
\label{eq:fx}
\end{align}
with $x_l = l + \frac{1}{2}$, is the effective size of the left block consisting of $l$ sites.
It is also known that the boundary oscillating component of the EE, $\mathcal{S}_{\rm b,osc}(l)$, is proportional to the oscillating part of the local energy density, $E_{\rm b,osc}(l)$, as\cite{LaflorencieSCA2006,AffleckLS2009}
\begin{align}
\mathcal{S}_{\rm b,osc}(l) = \alpha_{\rm b} E_{\rm b,osc}(l),
\label{eq:EE_to_bondE}
\end{align}
where $E_{\rm b,osc}(l)$ is obtained as
\begin{align}
E_{\rm b,osc}(l) &= E_{\rm b}(l) - E_{\rm b,uni},
\label{eq:bndE_stg}
\end{align}
with $E_{\rm b,uni} = \frac{1}{4}[E_{\rm b}(\frac{L}{2}-1)+2E_{\rm b}(\frac{L}{2})+E_{\rm b}(\frac{L}{2}+1)]$.

We have performed the least-square fitting of $\mathcal{S}(l)$ and $E_{\rm b,osc}(l)$ to the form Eq.\ (\ref{eq:EE_1Dform}) with Eq.\ (\ref{eq:EE_to_bondE}) by taking $c$, $\alpha_{\rm b}$, and $\mathcal{S}_0$ as fitting parameters.
We can thereby estimate the central charge $c$ from the DMRG data.
In Figs.\ \ref{fig:EEnt_Neel-Dim}(a) and \ref{fig:EEnt_Neel-Dim}(b), we show the EE $\mathcal{S}(l)$ and its uniform part
\begin{align}
\mathcal{S}_{\rm uni}(l) &= \mathcal{S}(l) - \alpha_{\rm b} E_{\rm b,osc}(l),
\label{eq:EEuni}
\end{align}
at the transition point between the N\'{e}el and dimer phases for $\gamma=1.0$.
It is clear in Fig.\ \ref{fig:EEnt_Neel-Dim}(b) that the uniform part of EE, $\mathcal{S}_{\rm uni}(l)$, for several system sizes collapses onto the same straight line when plotted as a function of $\ln[f(x_l)]$, suggesting that the fitting works well.
We have performed the same fitting procedure for the N\'{e}el-dimer transition points with various $\gamma$'s.
The central charge $c$ thus obtained is shown in Fig.\ \ref{fig:EEnt_Neel-Dim}(c).
The result clearly indicates that the central charge $c$ at large $L$ converges to the value $c=1$ for the entire transition line between N\'{e}el and dimer phases.
Therefore, we conclude that the N\'{e}el-dimer transition is the Gaussian transition with $c=1$.

\begin{figure}
\centering
\includegraphics[width=75mm]{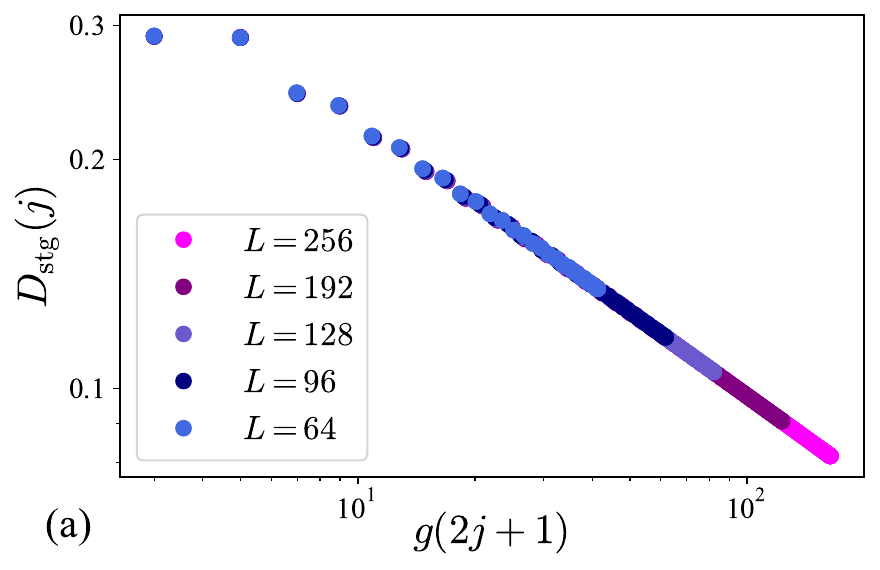}
\includegraphics[width=75mm]{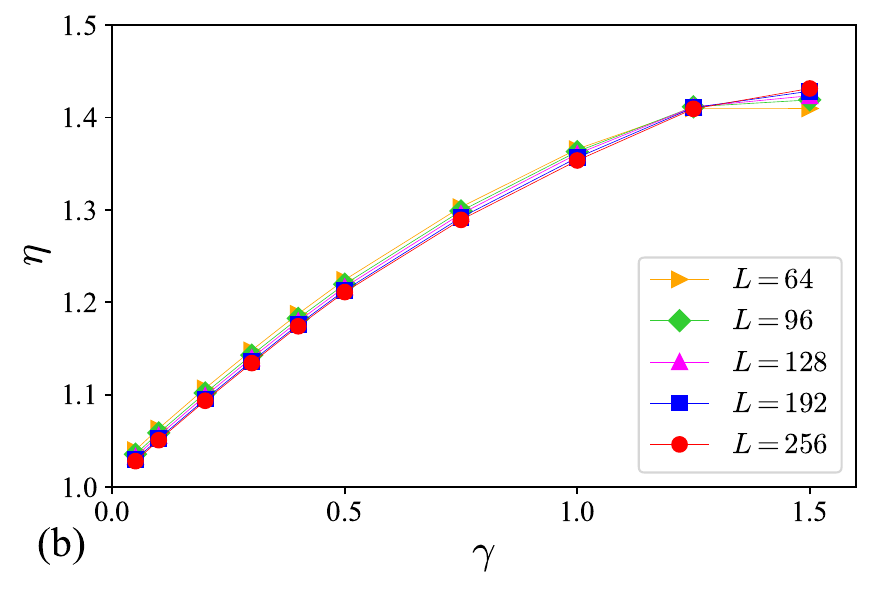}
\caption{
(a) Staggered part of the dimer operator, $D_{\rm stg}(j)$ as a function of $g(2j+1)$ at the N\'{e}el-dimer transition point for $\gamma=1.0$.
(b) TLL parameter $\eta$ at the N\'{e}el-dimer transition as a function of $\gamma$.
}
\label{fig:DimOpr_Neel-Dim}
\end{figure}

Finally, we examine the spatial profile of the dimer operator ${\bm S}_j \cdot {\bm S}_{j+1}$ at the N\'{e}el-dimer transition to estimate the TLL parameter.
At the Gaussian criticality on the N\'{e}el-dimer transition point, the expectation value of the dimer operator,
\begin{align}
D(j)=\langle {\bm S}_j \cdot {\bm S}_{j+1}\rangle,
\end{align}
in the system under the open boundary conditions takes the form,\cite{HikiharaFL2017}
\begin{align}
D(j) &\sim d_{\rm u} 
+ \frac{(-1)^j d_{\rm s}}{[g(2j+1)]^{\frac{1}{2\eta}}},
\label{eq:DimerOpr_fx_Neel-dim}
\end{align}
where we omit the higher-order terms.
The function $g(x)$ is defined as
\begin{align}
g(x) &= \frac{2(L+1)}{\pi} \sin\left(\frac{\pi |x|}{2(L+1)}\right).
\label{eq:gx}
\end{align}
We obtain the staggered part of the dimer operator as
\begin{align}
D_{\rm stg}(j) = (-1)^j \left[ D(j) - d_u \right]
\label{eq:Dstg}
\end{align}
with $d_u = \frac{1}{4}[D(\frac{L}{2}-1)+2D(\frac{L}{2})+D(\frac{L}{2}+1)]$.
Figure\ \ref{fig:DimOpr_Neel-Dim}(a) shows the data of $D_{\rm stg}(j)$ at the N\'{e}el-dimer transition point for $\gamma=1.0$, plotted as a function of $g(2j+1)$.
We find that $D_{\rm stg}(j)$ for several system sizes lies on the same straight line in the log-log plot, indicating that $D(j)$ satisfies the form (\ref{eq:DimerOpr_fx_Neel-dim}).
Thus, we can estimate the TLL parameter $\eta$ from the slope of the log-log plot of $D_{\rm stg}(j)$.
We have performed the same analysis along the N\'{e}el-dimer transition line by varying $\gamma$.
The TLL parameter $\eta$ thus obtained is shown in Fig.\ \ref{fig:DimOpr_Neel-Dim}(b).
It is found that when $\gamma$ increases from $\gamma=0$, $\eta$ rises from $\eta =1$, which is the value for the SU(2) symmetric case;
then, $\eta$ increases monotonically with $\gamma$.

\subsection{Transition between N\'{e}el and Ferromagnetic phases}\label{subsec:Neel-FM}

In this section, we discuss the phase transition between the N\'{e}el and FM phases that occurs in the parameter region where $J_2/J_1$ is small and $\gamma$ is large.
This transition was previously reported to be first order in Refs.\ \cite{SaitoH2024PRL,SaitoH2024PRB}.
Here, we show that the transition is a continuous one described by the Gaussian theory with the central charge $c=1$.

The low-energy effective theory describing this N\'{e}el-FM transition has been studied for the 1D anisotropic $J_1$-$J_2$ XZ model\cite{JiangM2019}.
We apply the theory to our model (\ref{eq:Ham_XYZ}), taking the parameter point $J_2/J_1 = 0$ and $\gamma \to \infty$ as the starting point.
Around the parameter point, we can rewrite the model (\ref{eq:Ham_XYZ}) as
\begin{align}
\bar{\mathcal{H}} &\equiv \frac{\mathcal{H}}{\gamma}
= \bar{\mathcal{H}}_0 + \bar{\mathcal{H}}',
\label{eq:Ham_Neel-FM} \\
\bar{\mathcal{H}}_0
&= J_1 \sum_j \left[
\bar{S}^x_j \bar{S}^x_{j+1} + \bar{S}^y_j \bar{S}^y_{j+1}
- \delta \bar{S}^z_j \bar{S}^z_{j+1} \right.
\nonumber \\
&\left.~~~+ \alpha \delta ( \bar{S}^x_j \bar{S}^x_{j+2} 
+ \bar{S}^y_j \bar{S}^y_{j+2} + \bar{S}^z_j \bar{S}^z_{j+2} ) \right],
\label{eq:Ham_Neel-FM_0} \\
\bar{\mathcal{H}}'
&= J_1 \sum_j \left[ 
- \delta ( \bar{S}^x_j \bar{S}^x_{j+1} - \bar{S}^y_j \bar{S}^y_{j+1} )
\right.
\nonumber \\
&\left.~~~
- \alpha ( \bar{S}^x_j \bar{S}^x_{j+2} - \bar{S}^y_j \bar{S}^y_{j+2} )
\right],
\label{eq:Ham_Neel-FM_p}
\end{align}
where $\alpha \equiv J_2/J_1 \ll 1$ and $\delta \equiv \frac{1}{\gamma} \ll 1$.
Here, we have introduced the spin-1/2 operators $\bar{\bm S}_j$, that are obtained by applying a $\pi$ rotation about $S^y$ axis to the ${\bm S}_j$ operators at every other site,
\begin{align}
\bar{S}^x_j = (-1)^j S^x_j,~~~
\bar{S}^y_j = S^y_j,~~~
\bar{S}^z_j = (-1)^j S^z_j.
\label{eq:StobarS}
\end{align}
Then, by applying the standard bosonization procedure to the Hamiltonian (\ref{eq:Ham_Neel-FM}), we obtain the effective Hamiltonian in terms of the bosonic fields $\{ \bar{\phi}, \bar{\theta} \}$ as
\begin{align}
\bar{\mathcal{H}} &= 
\frac{\bar{v}}{2} \int dx \left[
\frac{1}{\bar{\eta}} (\partial_x \bar{\theta})^2 
+ \bar{\eta} (\partial_x \bar{\phi})^2 \right]
\nonumber \\
&~~+ J_1 \bar{b}_2^2 (\delta - \alpha) \int dx \cos(\sqrt{8\pi} \bar{\theta}),
\label{eq:Ham_Neel-FM_boson}
\end{align}
where we have retained only the most relevant coupling term governing the N\'{e}el-FM transition.
The spin operators $\bar{\bm S}_j$ are related to the bosonic fields as
\begin{align}
\bar{S}^z_j &\simeq \frac{1}{\sqrt{2\pi}} \partial_x \bar{\phi}(x) 
+ \bar{b}_1 (-1)^j \sin(\sqrt{2\pi}\bar{\phi}(x)) + \cdots, 
\label{eq:barSz_to_bosonfield} \\
\bar{S}^+_j &\simeq e^{i\sqrt{2\pi}\bar{\theta}(x)}
\left[ \bar{b}_2 (-1)^j + \bar{b}_3 \sin(\sqrt{2\pi}\bar{\phi}(x))\right] 
+ \cdots.
\label{eq:barS+_to_bosonfield}
\end{align}
The coupling term of $\cos(\sqrt{8\pi} \bar{\theta})$ in the effective Hamiltonian (\ref{eq:Ham_Neel-FM_boson}) comes from $\bar{\mathcal{H}}'$ [Eq.\ (\ref{eq:Ham_Neel-FM_p})], while the Gaussian part comes from $\bar{\mathcal{H}}_0$ [Eq.\ (\ref{eq:Ham_Neel-FM_0})].
For $\alpha \ll 1$ and $\delta \ll 1$, $\bar{\mathcal{H}}_0$ is close to the 1D isotropic XY model, hence $\bar{\eta} \simeq \frac{1}{2}$, implying that the coupling term $\cos(\sqrt{8\pi} \bar{\theta})$ is relevant.
Therefore, when $\delta \ne \alpha$, the field $\bar{\theta}$ is pinned at the bottom of the cosine term.
The pinned value depends on the sign of $\delta-\alpha$;
$\bar{\theta} = \sqrt{\frac{\pi}{8}}$ or $3\sqrt{\frac{\pi}{8}}$ (mod $\sqrt{2\pi}$) for $\delta > \alpha$, and $\bar{\theta} = 0$ or $\sqrt{\frac{\pi}{2}}$ (mod $\sqrt{2\pi}$) for $\delta < \alpha$.
Thus, through the relations between the spin operators and the bosonic fields, $S^x_j = (-1)^j \bar{S}^x_j \sim \bar{b}_2 \cos(\sqrt{2\pi}\bar{\theta})$ and $S^y_j = \bar{S}^y_j \sim \bar{b}_2 (-1)^j \sin(\sqrt{2\pi}\bar{\theta})$, we arrive at the following conclusion;
The model exhibits the N\'{e}el phase with the magnetic order
\begin{align}
\langle S^x_j \rangle = 0,~~~
\langle S^y_j \rangle = \pm (-1)^j c^y,
\end{align}
for $\delta > \alpha$, and the FM phase with
\begin{align}
\langle S^x_j \rangle = \pm c^x,~~~
\langle S^y_j \rangle =0,
\end{align}
for $\delta < \alpha$, where $c^x$ and $c^y$ are constants.
At $\delta=\alpha$, the model undergoes the phase transition with a criticality governed by the Gaussian model.

\begin{figure}
\centering
\includegraphics[width=80mm]{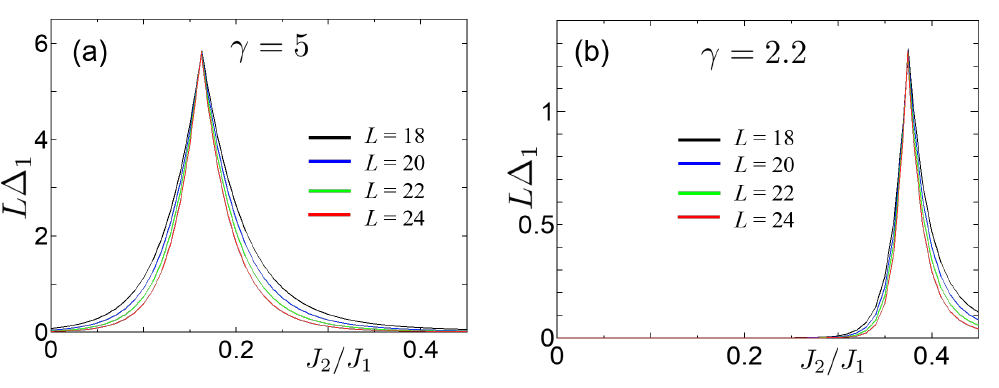}
\caption{
Scaled gap $L\Delta_1(L)$ as a function of $J_2/J_1$ along parameter lines with (a) $\gamma=5.0$ and (b) $\gamma=2.2$.
}
\label{fig:SGap_Neel-FM}
\end{figure}

To verify the validity of the effective theory and elucidate the nature of the phase transition, we have performed numerical analyses of the N\'{e}el-FM transition, as in the case of the N\'{e}el-dimer transition discussed in Sec.\ \ref{subsec:Neel-Dim}.
First, we have examined the scaled gap to determine the transition points.
In both the N\'{e}el and FM phases, the system possesses the doubly-degenerate ground states in the thermodynamic limit, and the energy levels of the first-excited states of the finite periodic chain in the two phases cross at the transition.
Therefore, the boundary between the N\'{e}el and FM phases can be identified by the cusp singularity in the first excitation gap $\Delta_1(L)$ of the finite system under the periodic boundary conditions.
In Fig.\ \ref{fig:SGap_Neel-FM}, we present the data of the scaled gap $L \Delta_1(L)$ obtained by the ED method for the parameter line with $\gamma=5.0$ and $2.2$.
It is clearly seen in the figure that the scaled gap exhibits a cusp singularity.
On both sides of the cusp, the scaled gap decreases with $L$, indicating that the gap $\Delta_1(L)$ decreases faster than $1/L$, presumably exponentially.
The height of the cusp in the scaled gap $L \Delta_1(L)$ is almost independent of the system size $L$, suggesting that the gap $\Delta_1(L)$ scales as $\sim 1/L$ at the transition point.
This observation is consistent with the prediction of the effective theory that the transition is continuous and described by the Gaussian theory.
For various $\gamma$, we have estimated the critical value of $J_2/J_1$ at the cusp of the scaled gap for finite-size systems.
We have then extrapolated the values by fitting them to a quadratic polynomial of $1/L$ to determine the N\'{e}el-FM transition points, which are plotted in Fig.\ \ref{fig:GSPD}.

\begin{figure}
\centering
\includegraphics[width=75mm]{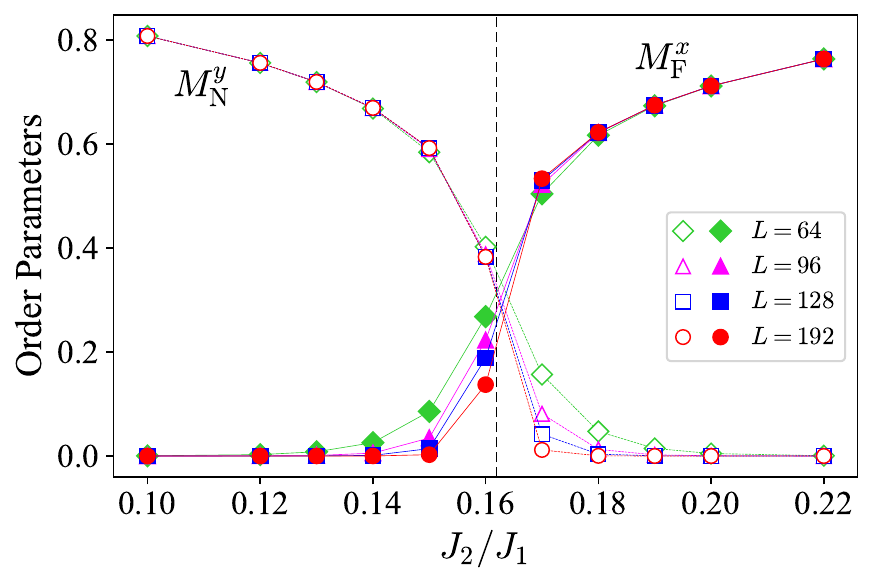}
\caption{
Order parameters as a function of $J_2/J_1$ along the parameter line with $\gamma=5.0$.
Open and solid symbols represent the N\'{e}el order parameter $M^y_{\rm N}$ and the ferromagnetic order parameter $M^x_{\rm F}$, respectively.
Vertical dashed line represents the boundary between the N\'{e}el and FM phases.
}
\label{fig:OP_Neel-FM}
\end{figure}

To confirm the long-range order characterizing the phase on either side of the transition point, we have calculated the order parameters $M^y_{\rm N}$ and $M^x_{\rm F}$ using the DMRG method.
The calculation was performed for the system under the open boundary conditions with the boundary fields $h^x_1 = h^x_L = h^y_1 = -h^y_L = 100J_1/\sqrt{2}$, which lift the degeneracy of the ground states in both the N\'{e}el and FM phases.
Figure\ \ref{fig:OP_Neel-FM} shows the results for the parameter line with $\gamma = 5.0$.
We find in the figure that the model exhibits the N\'{e}el long-range order along the $S^y$ direction in the N\'{e}el phase for small $J_2/J_1$, and the FM long-range order along the $S^x$ direction in the FM phase for large $J_2/J_1$.
The result confirms that the transition is indeed the N\'{e}el-FM one.
The order parameters decrease continuously as $J_2/J_1$ approaches the transition point, indicating that the transition is continuous.

\begin{figure}
\centering
\includegraphics[width=75mm]{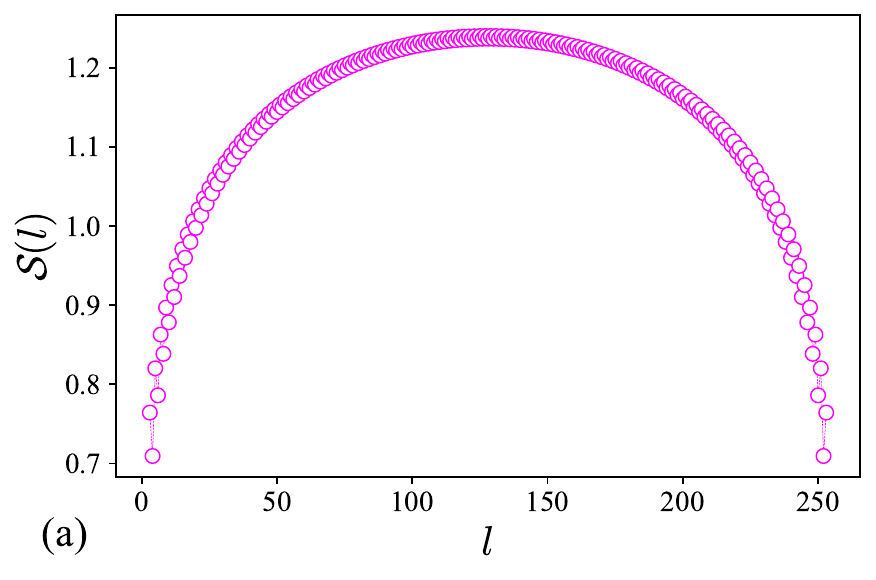}
\includegraphics[width=75mm]{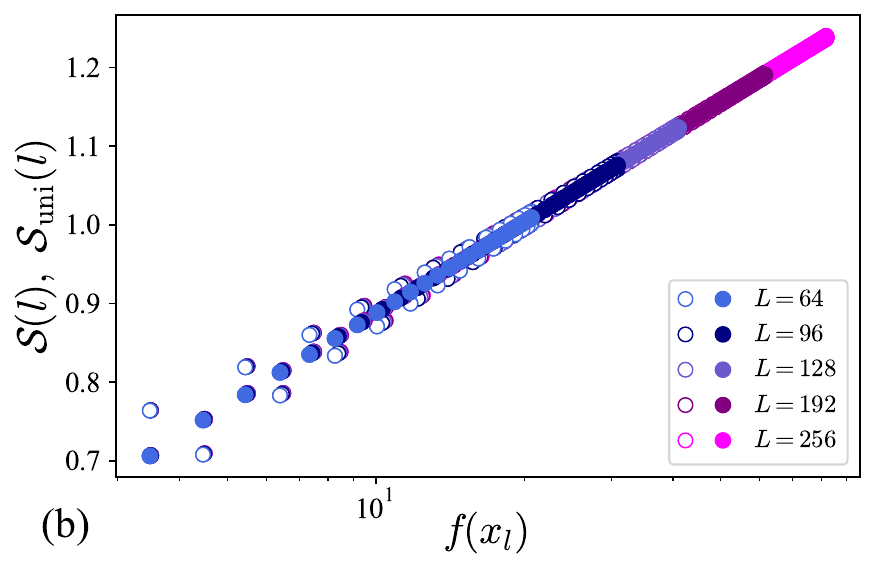}
\includegraphics[width=75mm]{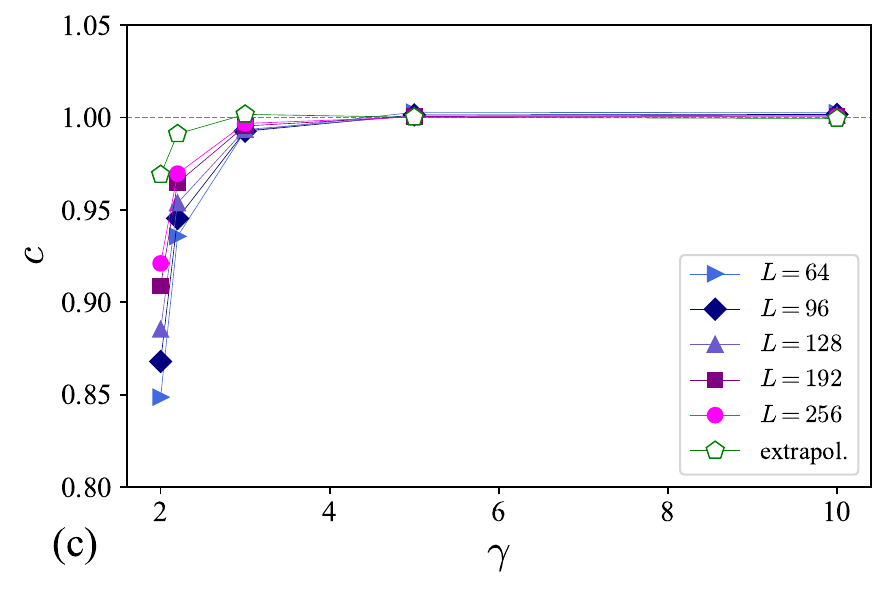}
\caption{
(a) Entanglement entropy $\mathcal{S}(l)$ of the system with $L=256$ spins at the N\'{e}el-FM transition point for $\gamma=5.0$, plotted as a function of $l$.
(b) $\mathcal{S}(l)$ (open symbols) and its uniform part $\mathcal{S}_{\rm uni}(l)$ (solid symbols) for several $L$ at the N\'{e}el-FM transition point for $\gamma=5.0$, plotted as a function of $f(x_l)$.
(c) The central charge $c$ at the N\'{e}el-FM transition as a function of $\gamma$ for several $L$.
The green pentagons represent the extrapolated values.\cite{extrapl_footnote}
}
\label{fig:EEnt_Neel-FM}
\end{figure}

Next, to explore the nature of the transition, we have performed the DMRG calculation of the EE $\mathcal{S}(l)$ and the local energy density $E_{\rm b}(l)$ along the transition line.
No boundary field is applied in this calculation since no ground-state degeneracy is expected at the transition points.
Figures\ \ref{fig:EEnt_Neel-FM}(a) and \ref{fig:EEnt_Neel-FM}(b) show the DMRG data of the EE $\mathcal{S}(l)$ and its uniform part $\mathcal{S}_{\rm uni}(l)$.
By fitting the data of $\mathcal{S}(l)$ and $E_{\rm b,osc}(l)$
to the form Eq.\ (\ref{eq:EE_1Dform}) with Eq.\ (\ref{eq:EE_to_bondE}), we have estimated the central charge $c$.
The estimates of $c$ at the transition points with different $\gamma$ are presented in Fig.\ \ref{fig:EEnt_Neel-FM}(c).
It is found that the estimates for $\gamma \ge 3.0$ are very close to unity, $c=1$.
For $\gamma=2.20$ and $2.00$, where the transition points are in the vicinity of the multicritical point, the size dependence of $c$ is noticeable.
However, the fitting of $c(L)$ 
to the form $c(L) = c(\infty) + \alpha_{c1}/L + \alpha_{c2}/L^2$ yields the extrapolated value $c(\infty)$ close to unity, $c(\infty)= 0.991$ for $\gamma=2.20$ and $c(\infty) = 0.969$ for $\gamma=2.00$.\cite{extrapl_footnote}
Our numerical result thus suggests that the central charge takes the universal value $c=1$ at the N\'{e}el-FM transition, although the precise estimation of $c$ near the multicritical point would be left for future work.

\begin{figure}
\centering
\includegraphics[width=75mm]{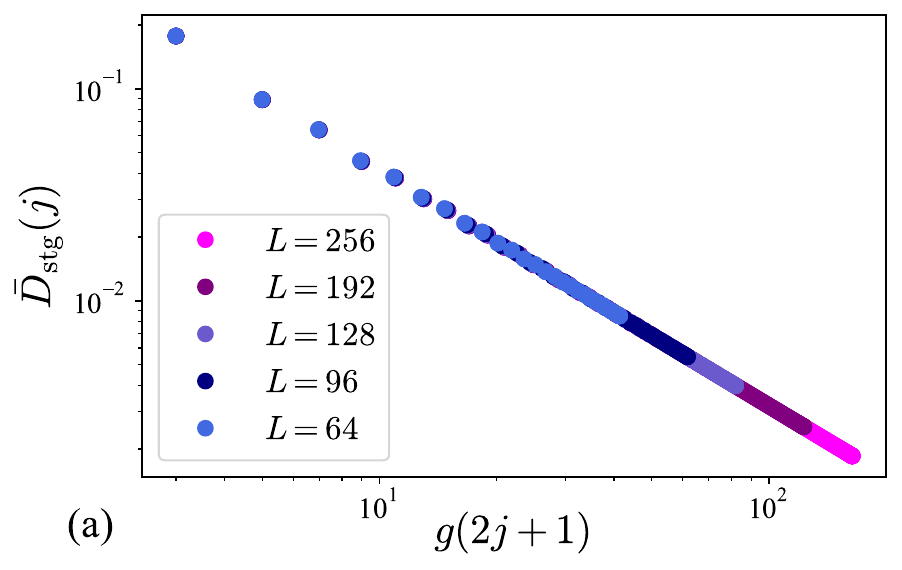}
\includegraphics[width=75mm]{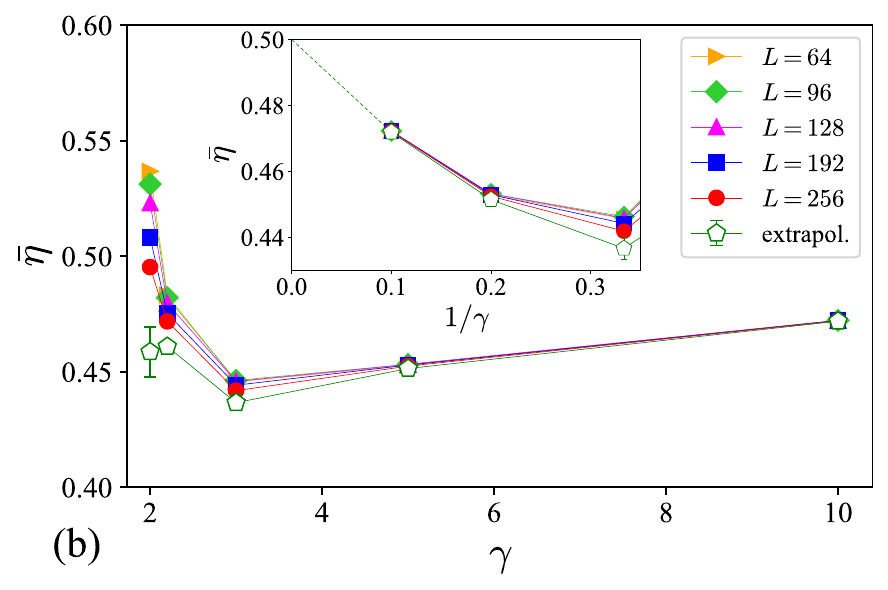}
\caption{
(a) Staggered part of the dimer operator, $\bar{D}_{\rm stg}(j)$, as a function of $g(2j+1)$ at the N\'{e}el-FM transition point for $\gamma=5.0$.
(b) TLL parameter $\bar{\eta}$ at the N\'{e}el-FM transition as a function of $\gamma$ for several $L$.
The green pentagons represent the extrapolated values.\cite{extrapl_footnote}.
Inset shows $\bar{\eta}$ for $\gamma \ge 3.0$ as a function of $1/\gamma$.
}
\label{fig:DimOpr_Neel-FM}
\end{figure}

Finally, we determine the TLL parameter at the N\'{e}el-FM transition from the dimer operator.
Considering the transformation in Eq.\ (\ref{eq:StobarS}), we analyze the following dimer operator,
\begin{align}
\bar{D}(j) = \langle \bar{\bm S}_j \cdot \bar{\bm S}_{j+1} \rangle
= \langle (- S^x_j S^x_{j+1} + S^y_j S^y_{j+1} - S^z_j S^z_{j+1} )\rangle.
\label{eq:barD}
\end{align}
At the transition, $\bar{D}(j)$ is represented by Eq.\ (\ref{eq:DimerOpr_fx_Neel-dim}) with $\eta$ replaced by $\bar{\eta}$.
Figure\ \ref{fig:DimOpr_Neel-FM}(a) shows the staggered part of the dimer operator, $\bar{D}_{\rm stg}(j)$, which is obtained in the same manner as in Eq.\ (\ref{eq:Dstg}), at the N\'{e}el-FM transition point for $\gamma=5.0$.
We see in the figure that the data of $\bar{D}_{\rm stg}(j)$ for various system sizes collapse onto a single straight line in the log-log plot;
therefore, we can determine the TLL parameter $\bar{\eta}$ from the slope of the log-log plot.
The estimates of $\bar{\eta}$ thus obtained for various $\gamma$ are shown in Fig.\ \ref{fig:DimOpr_Neel-FM}(b).
The extrapolation to $L \to \infty$ was performed in the same way as that for the central charge.\cite{extrapl_footnote}
In the figure, we find that $\bar{\eta}$ asymptotically approaches $\frac{1}{2}$ as $\gamma \to \infty$, as expected from the fact that the system becomes the isotropic XY model at $\gamma \to \infty$.
$\bar{\eta}$ also appears to approach $\frac{1}{2}$ as $\gamma \to \sqrt{3}^+$, the value of $\gamma$ at the multicritical point.
However, the limited accuracy of the estimates near the multicritical point prevents us from confirming the behavior.
Clarifying this issue remains a subject for future study.

\subsection{Transitions to Double-N\'{e}el phase}\label{subsec:DN}

In the ground-state phase diagram of the model (\ref{eq:Ham_gamma}) or (\ref{eq:Ham_XYZ}), the DN phase appears in the region for large $J_2/J_1$ and intermediate $\gamma$.
The DN phase is characterized by the magnetic long-range order of the up-up-down-down pattern along the $S^y$ direction, and has the four-fold degenerate ground states in the thermodynamic limit.
The DN phase is adjacent to the dimer and FM phases, and the dimer-DN and FM-DN transition lines converge to the multicritical point, where the N\'{e}el-dimer and N\'{e}el-FM transition lines also meet.

\begin{figure}
\centering
\includegraphics[width=75mm]{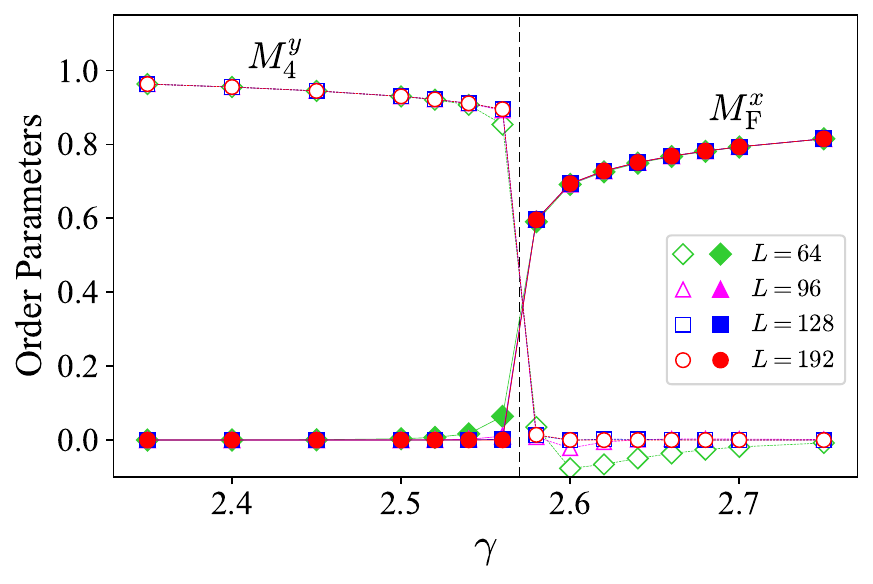}
\caption{
Order parameters as a function of $\gamma$ along the parameter line with $J_2/J_1=1.0$.
Open and solid symbols represent the double-N\'{e}el order parameter $M^y_4$ and the ferromagnetic order parameter $M^x_{\rm F}$, respectively.
Vertical dashed line represents the boundary between the DN and FM phases.
}
\label{fig:OP_FM-DN}
\end{figure}

First, we discuss the transition between the FM and DN phases.
Figure\ \ref{fig:OP_FM-DN} presents the order parameters $M^x_{\rm F}$ and $M^y_4$ on the parameter line with $J_2/J_1=1.0$ around the FM-DN transition point.
The data were obtained by the DMRG calculation for the system under the open boundary conditions with the boundary fields $h^x_1 = h^x_L = h^y_1 = h^y_L = 100J_1/\sqrt{2}$, which lift the degeneracy of the ground states in both phases.\cite{hbound_FM-DN_footnote}
The size dependence of the data is nearly negligible at $L=192$.
It is also found that both the order parameters exhibit a finite jump at the transition point.
The result suggests that the transition is first order.
The FM-DN transition points determined in this manner are plotted in the phase diagram, Fig.\ \ref{fig:GSPD}.

Next, we discuss the transition between the dimer and DN phases.
To determine the phase boundary between the two phases, we have calculated the low-lying energy levels by the ED method for systems with up to $L=28$ spins under the periodic boundary conditions.
In the thermodynamic limit, the ground states are two-fold degenerate in the dimer phase and four-fold degenerate in the DN phase.
Therefore, in the dimer phase, the level splitting between the ground and first-excited states in the finite-size systems, $\Delta_1(L)$, decays rapidly with increasing the system size $L$, while the splitting to the second-excited state, $\Delta_2(L)$, remains finite and converges to the finite excitation gap above the doubly-degenerate ground states as $L \to \infty$.
In contrast, in the DN phase, the four lowest-energy states become degenerate and $\Delta_n(L)$ with $n=1,2,3$ decay to zero as $L$ increases.
To detect this difference, we monitor the system-size dependence of the scaled gap $L\Delta_2(L)$;
it is expected that $L\Delta_2(L)$ increases linearly with $L$ in the dimer phase, while it decreases rapidly (presumably exponentially) with $L$ in the DN phase.

\begin{figure}
\centering
\includegraphics[width=80mm]{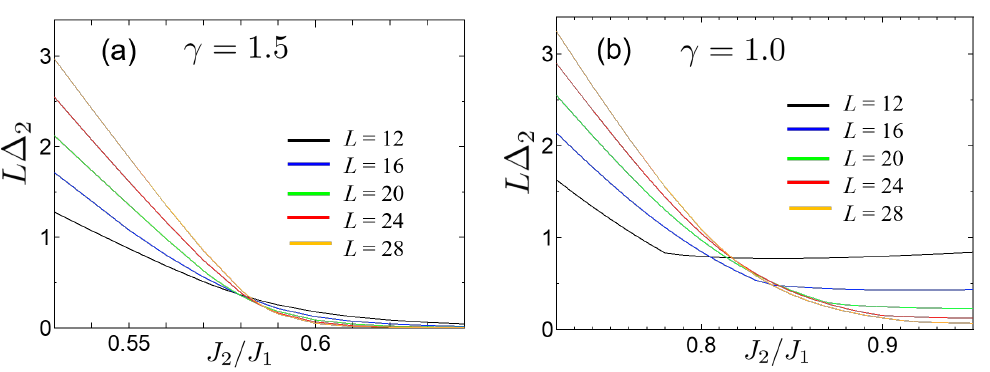}
\caption{
Scaled gap $L\Delta_2(L)$ as a function of $J_2/J_1$ along parameter lines with (a) $\gamma=1.5$ and (b) $\gamma=1.0$.
}
\label{fig:ScaledGap-Dimer-DN}
\end{figure}

Figure~\ref{fig:ScaledGap-Dimer-DN} shows the ED data of $L\Delta_2(L)$ for $L = 12$ - $28$ as a function of $J_2/J_1$ for $\gamma=1.5$ and $1.0$.
The size dependence of $L\Delta_2(L)$ changes with $J_2/J_1$;
specifically, $L\Delta_2(L)$ increases with $L$ for small $J_2/J_1$, whereas it decreases with $L$ for large $J_2/J_1$.
These behaviors are consistent with those expected in the dimer and DN phases, respectively.
We have determined the critical value of $J_2/J_1$ for finite-size systems with up to $L=28$ spins using the crossing condition
\begin{align}
L \Delta_2(L) = (L-4) \Delta_2(L-4).
\end{align}
Here, we have found that the finite-size critical value often exhibits irregular $L$-dependence, presumably due to the incommensurate character of the excitations in the dimer phase, and the extrapolation to $L \to \infty$ does not work well.
Thus, we have adopted the critical value of $J_2/J_1$ for $L=28$ as our estimate of the transition point.
In this manner, we have estimated the transition points along parameter lines with fixed $\gamma$ or fixed $J_2/J_1$.
The results are plotted in Fig.\ \ref{fig:GSPD} with error bars determined from the difference between the estimates for $L=28$ and $L=24$.
We note that the estimated transition points may contain non-negligible errors.
The precise determination of the dimer-DN transition line and the elucidation of the nature of the transition are left for future study.

\section{Conclusion}\label{sec:conclusion}

In summary, we have investigated the ground state of the spin-1/2 Heisenberg zigzag ladder with the anisotropic $\Gamma$ term [Eq.\ (\ref{eq:Ham_gamma}) with Eq.\ (\ref{eq:def_gamma})], that is equivalent to the spin-1/2 XYZ zigzag ladder [Eq.\ (\ref{eq:Ham_XYZ})].
For the models, the ground-state phase diagram was proposed in Refs.\ \cite{SaitoH2024PRL,SaitoH2024PRB}.
In this paper, by combining analytical approaches such as the bosonization method and the numerical techniques such as the ED and the DMRG methods, we have revisited the phase diagram and updated it.
We have shown that in the parameter region of $\gamma > 0$, where the model is anisotropic, the model exhibits three magnetic ordered phases, the N\'{e}el, ferromagnetic, and double N\'{e}el phases, and the dimer phase.
These four phases have a finite excitation gap and meet at the multicritical point at $(J_2/J_1, \gamma)=(1/2, \sqrt{3})$.
The TLL phase appears only on the parameter line of $\gamma=0$, where the model possesses the SU(2) symmetry of spin rotation.
Furthermore, we have identified that the transitions between the N\'{e}el and dimer phases and between the N\'{e}el and ferromagnetic phases are both continuous transitions governed by the Gaussian criticality with the central charge $c=1$.
The TLL parameter along these Gaussian transition lines has also been estimated numerically.

The model (\ref{eq:Ham_gamma}) has been proposed to describe the Yb-based frustrated semiconductor YbCuS$_2$\cite{Ohmagari2020}.
The corresponding parameters are expected to be $J_1 \sim J_2$ and $\gamma = \Gamma_1/J_1 = \Gamma_2/J_2 \sim 10^{-1}$ - $10^{-2}$,\cite{SaitoNH2024} suggesting that YbCuS$_2$ could be in the gapped dimer phase with a slight anisotropy.
Furthermore, a high pressure experiment of the compound has suggested that the wave number of the magnetic order changes from incommensurate at ambient pressure to commensurate at 1.6 GPa, indicating that applying the pressure decreases $J_2/J_1$ to a value less than $0.5$.\cite{HoriMKISO2025}
Verifying whether the model (\ref{eq:Ham_gamma}) can comprehensively explain experimental findings remains an important issue for future research.
We are currently performing the calculations of finite-temperature properties of the model, and the results will be reported elsewhere.\cite{MaeshimaOH_unpub}

\begin{acknowledgments}
The authors thank A.\ Furusaki, H.\ Saito, and C.\ Hotta for fruiteful discussions and insightful comments.
This work is partially supported by KAKENHI Grant Numbers JP21H05182, JP21H05191, JP24K06881, JP24K06913, and JST CREST Grant Number JPMJCR24I1.
Numerical calculations were performed in part on the Pegasus System at the Center for Computational Sciences, University of Tsukuba, and the Miyabi System in the Joint Center for Advanced High Performance Computing.

\end{acknowledgments}

% Use a BibTeX database file, e.g., references.bib.
%\bibliography{AFzigzag-gamma_Ref}

\begin{thebibliography}{56}
\bibitem [{SSB()}]{SSB_footnote}%
  \BibitemOpen
  \href@noop {} {}\bibinfo {note} {In the bulk of a real compound, one of the
  degenerate ground states is realized due to the spontaneous symmetry
  breaking. This spontaneous symmetry breaking also occurs in the other ordered
  phases with degenerate ground states.}\BibitemShut {Stop}%
\bibitem [{ext()}]{extrapl_footnote}%
  \BibitemOpen
  \href@noop {} {}\bibinfo {note} {The extrapolated value was obtained by
  fitting the data for $L \ge 64$. The error in the extrapolated value was
  estimated from the absolute difference between this value and that obtained
  by fitting the data for $L \ge 96$. We note that in Fig.\
  \ref{fig:EEnt_Neel-FM}(c), the error was smaller than the size of the symbols
  (pentagons).}\BibitemShut {Stop}%
\bibitem [{hbo()}]{hbound_FM-DN_footnote}%
  \BibitemOpen
  \href@noop {} {}\bibinfo {note} {In the DN phase, the four-site-period
  magnetic order can start from each system edge in four different patterns,
  such as up-up-down-down-$\cdots$ and up-down-down-up-$\cdots$, depending on
  the model parameters $J_2/J_1$ and $\gamma$ and the boundary fields. Since
  the orders induced from the two edges by the boundary fields can be
  incompatible with each other in the bulk, we have also performed an analysis
  using the boundary fields $h^x_1 = h^x_L = h^y_1 = -h^y_L = 100J_1/\sqrt{2}$
  and the order parameter $M'^y_4 = \frac{1}{2} \langle ( S^y_{\frac{L}{2}-1} +
  S^y_{\frac{L}{2}} - S^y_{\frac{L}{2}+1} - S^y_{\frac{L}{2}+2} ) \rangle$, in
  addition to the analysis presented in the main text. The combined results of
  the two analyses indicate the first-order transition between the FM and DN
  phases.}\BibitemShut {Stop}%
\end{thebibliography}

\end{document}